\documentclass[pra,aps,twocolumn,epsfig,superscriptaddress,showpacs,showkeys]{revtex4}
\usepackage{amsmath}
\usepackage{amssymb}
\usepackage{graphicx}% Include figure files
\usepackage{dcolumn}% Align table columns on decimal point
\usepackage{bm}% bold math
\usepackage[colorlinks=false,dvipdfm]{hyperref} % citecolor=blue 文件内部引用
\usepackage{setspace}
\usepackage{amsfonts}
\usepackage{rotating}
\usepackage{makeidx}
\usepackage{amsthm}
\usepackage{amsmath}

\begin{document}
\title{Unambiguous state discrimination with intrinsic coherence\footnote{Entropy 2022, 24, 18. https://doi.org/10.3390/e24010018}}

\author{Jin-Hua Zhang}
\email[Corresponding author: ]{zhjh217@sina.com}
\affiliation{Department of
Physics, Xinzhou Teacher's University, Xinzhou 034000, China}

\author{Fu-Lin Zhang}
\email[Corresponding author: ]{flzhang@tju.edu.cn}
\affiliation{Department of Physics, School of Science, Tianjin
University, Tianjin 300072, China}

\author{Zhi-Xi Wang}
\affiliation{School of Mathematical Sciences, Capital Normal
University, Beijing 100048, China}

\author{Hui Yang}
\affiliation{Department of
Physics, Xinzhou Teacher's University, Xinzhou 034000, China}

\author{Shao-Ming Fei}
\email[Corresponding author: ]{feishm@cnu.edu.cn}
\affiliation{School of Mathematical Sciences, Capital Normal
University, Beijing 100048, China}
\affiliation{Max-Planck-Institute for Mathematics in the Sciences,
D-04103 Leipzig, Germany}

%\date{\today}

\begin{abstract}
We investigate the discrimination of pure-mixed (quantum filtering) and mixed-mixed states and compare their optimal success probability with the one for discriminating other pairs of pure states superposed by the vectors included in the mixed states. We prove that
under the equal-fidelity condition, the pure-pure state discrimination scheme is superior to the pure-mixed (mixed-mixed) one. With respect to quantum filtering, the coherence exists only in one pure state and is detrimental to the state discrimination for lower dimensional systems; while it is the opposite for the mixed-mixed case with symmetrically distributed coherence. Making an extension to infinite-dimensional systems, we find that the coherence which is detrimental to state discrimination may become helpful and vice versa.
\end{abstract}


% insert suggested PACS numbers in braces on next line
%\pacs{03.65.-w; 02.20.-a; 21.10.sf; 31.30.jx}
% insert suggested keywords - APS authors don't need to do this
%\keywords{}

%\maketitle must follow title, authors, abstract, \pacs, and \keywords
\maketitle
%\end{CJK*}

% body of paper here - Use proper section commands
% References should be done using the \cite, \ref, and \label commands
% \section{Introduction \label{intro}}
% Put \label in argument of \section for cross-referencing
%\section{\label{}}

%{\bf PACS numbers:} 03.67.-a, 03.65.Ta

%\section{Introduction}
%\emph{Introduction.--}

%<1>-----------entanglement and discord

%\noindent \emph{Introduction.---}

\section{Introduction} \label{intro}

Quantum state discrimination is of great importance in quantum information processing \cite{Enk2002PRA}. A fundamental result in quantum mechanics is the impossibility to
distinguish perfectly two or more non-orthogonal quantum states. It is then a key task to discriminate the states with maximal success probability. Such state discrimination problems
branch out into two important streams: ambiguous \cite{Wittmann2008,Tsujino2011,Assalini2011,Xiong2018JPA,Xiong2019PRA} and unambiguous quantum state discrimination \cite{Ivanovic1987PLA,Peres1988PLA,Dies1988PLA,Bennett1992PRL,Bergou2003PRL,Pang2009PRA,Pang2013PRA,Bergou2013PRL,Namkung2017PRA,Namkung2018SR,Zhang2017arXiv}.
The study on minimization of the error in the ambiguous state discrimination was pioneered by Helstrom who provided a lower bound on the error probability in distinguishing two quantum states. This bound can be attained through the ways presented in \cite{Wittmann2008,Tsujino2011,Assalini2011}.  While the unambiguous quantum state discrimination is \mbox{error-free \cite{Ivanovic1987PLA,Peres1988PLA,Dies1988PLA,Bennett1992PRL,Bergou2003PRL,Pang2009PRA}}. It plays key roles in various contexts in quantum information theory, including quantum key distribution \cite{Bergou2003PRL,Bergou2013PRL,Namkung2018SR}, the study of quantum correlations \cite{Roa2011PRL,Li2012PRA,Zhang2013SR,Jimenez}, and the role of entanglement in local discrimination of bipartite systems \cite{JHZhang2020PRA}.

%Just like quantum correlation and entanglement, quantum coherence is also originated from the quantum superposition principle and fundamental in quantum physics \cite{Aberg2014PRL,Levi2014NJP}.

%\deleted{Quantum correlations such as quantum entanglement is tightly related to quantum coherence} \added{
Quantum coherence is also a critical resource in quantum state discrimination and is tightly related to quantum correlations such as quantum entanglement \cite{zhuhj2017PRA}.  Recently, the quantification of quantum coherence has been extensively studied in the framework of quantum resource theory \cite{Aberg2014PRL,Levi2014NJP,bukf,jinzx}. The role of coherence played in ambiguous state discriminations \cite{Xiong2018JPA,Xiong2019PRA} has been investigated. There are also a few results on unambiguous state discriminations with coherence which is generated or consumed in auxiliary systems and utilized as resources \cite{Kim2018arXiv}. Actually, the coherence in \cite{Kim2018arXiv} comes from the non-orthogonality of the initial states.

%It is interesting to uncover the roles of quantum coherence contained in a composite quantum system in the procedure of state discrimination. Recently, for ambiguous state discrimination, the analytic relation between minimum failure probability and coherence which is dependent on the wave-particle dualism is founded in \cite{Xiong2018JPA,Xiong2019PRA}. For unambiguous state discrimination assisted by an auxiliary system, it is found that quantum coherence generated or consumed in the auxiliary system can be also utilized as a resource \cite{Kim2018arXiv}. Since this coherence reflects the non orthogonality of the initial states, it is positively correlated with the optimal success probability as a mater of course.

{In this work,} different from the results in \cite{Kim2018arXiv}, we consider the effect of the coherence encoded in the initial state on unambiguous state discriminations. We first apply a quantum state filtering \cite{Bergou2005PRA}, which is the discrimination between
a pure state from another rank-$N$ incoherent mixed state composed of $N$ vectors.
Then, we superpose these $N$ vectors into a new pure state and then do a pure-pure state discrimination. If the fidelity of the pure-pure state equals the pure-mixed one, it can be proved that the pure-pure scheme is superior to the pure-mixed one; but the coherence is detrimental to the state discrimination for lower dimensional systems.  Furthermore, through the discrimination of two rank-$N$ mixed states and the comparison with the results of another pure-pure-state discrimination scheme,  as an extension of the results in \cite{JHZhang2020PRA}, we prove that pure-pure scheme is still superior to mixed-mixed one if the eigenvectors of the mixed states have a one-to-one overlap (an equal-fidelity case); but there exists a great deal of symmetrically distributed coherence which is helpful to state discrimination, in contrary to the result of quantum filtering.

Finally, we extend
%\deleted{from finite} \added{
the results to infinite-dimensional systems where the vectors included in the mixed states are mixed with each other via the probability factors coinciding with the photon number distribution of two kinds of Gaussian states in quantum \mbox{optics \cite{YRZhang2016PRA}.}
 We find that corresponding to the well-known coherent state, the symmetrically (asymmetrically) distributed coherence may become detrimental (helpful) on the contrary,
which can be attributed to the fact that the well-known coherent state approaches the boundary between classical and quantum physics.

The paper is organized as follows. In Section \ref{mixed local state
discrimination1}, we present the result of quantum state filtering. Additionally, we compare its results with the one of the pure-pure state schemes. In Section \ref{Simulation1}, we compare the discrimination of two rank-$N$ mixed states with the scheme for discriminating other two pure states having the same fidelity with the mixed ones. We generalize the results to infinite systems associated with two kinds of Gaussian states in Section \ref{infinite dimensional}. We summarize in the last section.

\section{Quantum state filtering}\label{mixed local state discrimination1}

Consider a set of given $N+1$ nonorthogonal quantum states $\{|\Psi_1\rangle,|\Psi_{i'}\rangle\}$ ($i=1,2...,N$), occurring with prior probability $P_1, P_2\beta_i$, where $P_1+P_2=1$ and $\sum\limits_{i=1}^N\beta_i=1$, $\beta_i\geq0$.
We want to find a procedure that unambiguously assigns the state of
the quantum system to one or the other of two complementary subsets of the set of the $N+1$ given nonorthogonal quantum states, namely, either $|\Psi_1\rangle$ or $\{|\Psi_{i'}\rangle\}$. This filtering problem (pure-mixed state discrimination) \cite{Bergou2005PRA} is equivalent to the problem of discrimination between a pure (rank $1$) state $\rho_1$ and an arbitrary (rank-$N$) mixed state $\rho_2$,
\begin{eqnarray}
\rho_1&=&|\Psi_1\rangle\langle\Psi_1|,\nonumber\\
\rho_2&=&\sum\limits_{i=1}^N\beta_i|\Psi_{i'}\rangle\langle\Psi_{i'}|,
\end{eqnarray}
prepared with the prior probability $P_1$ and $P_2$ ($P_1\leq P_2$). For simplicity, we assume that the following relations are fulfilled:
\begin{equation}\label{overlap}
\langle\Psi_1|\Psi_{i'}\rangle=s_{1i'}\geq0, ~~ \langle\Psi_{i'}|\Psi_{j'}\rangle=\delta_{ij}
\end{equation}
for $i,j=1,...,N$. Let $|\Psi^{\parallel}_1\rangle$ be the component of $|\Psi_1\rangle$ in the subspace spanned by the vectors $|\Psi_{1'}\rangle, |\Psi_{2'}\rangle,...,|\Psi_{n'}\rangle$. We have
\begin{equation}\label{inner product inequality}
\langle \Psi^{\parallel}_1|\Psi^{\parallel}_1\rangle=\sum\limits_{i=1}^Ns_{1i'}^2<1.
\end{equation}

In order to discriminate the two sets unambiguously, we couple the system with an ancilla $|k_a\rangle$ \cite{Zhang2017arXiv,Pang2013PRA,Bergou2005PRA} via tensor product method \cite{Bergou2005PRA,Chen2007PRA} and perform a joint unitary transformation $U$,
\begin{eqnarray}\label{unitary evolution 1}
U|\Psi_1\rangle|k_a\rangle\!\!&=&\!\!\sqrt{q_1}e^{i\gamma_1}|\phi_0\rangle|0\rangle_a+\sqrt{1-q_1}|\Phi_1\rangle|1\rangle_a,\nonumber\\
U|\Psi_{i'}\rangle|k_a\rangle\!\!&=&\!\!\sqrt{q_{i'}}e^{i\gamma_{i'}}|\phi_0\rangle|0\rangle_a+\sqrt{1-q_{i'}}|\Phi_{i'}\rangle|1\rangle_a.
\end{eqnarray}
Since we are aiming to discriminate $|\Psi_1\rangle$ from $|\Psi_{i'}\rangle$ optimally,
it is required that the post-measured state $|\Phi_1\rangle$ is orthogonal to $|\Phi_{i'}\rangle$,
$\langle\Phi_1|\Phi_{i'}\rangle=0$ for $i=1,2,...,N$, while $\langle \Phi_{i'}|\Phi_{j'}\rangle\neq0$ for $i\neq j$, $i,j=1,2,...,N$.

Thus, after a
von-Neumann measurement on the ancilla, the vector $|\Psi_1\rangle$ is distinguished from the set $\{|\Psi_{i'}\rangle\}$ successfully if the measurement outcome is $|1\rangle_a$, otherwise the outcome $|0\rangle_a$ implies failure. The average failure probability $Q$ is given by
\begin{equation}
Q=P_1q_1+\sum\limits_{i=1}^N P_2\beta_iq_{i'},
\end{equation}
where the parameters $q_1$ and $q_i$ satisfy $q_1q_i=s_{1i'}^2$ according to Eq. (\ref{unitary evolution 1}). Therefore, the optimization of $Q$ is given by
\begin{equation}
{\rm{minimize}}\ Q=P_1q_1+\frac{\sum\limits_{i=1}^N P_2\beta_is_{1i'}^2}{q_1},
\end{equation}
\begin{equation}\label{inner product}
{\rm{subject\ to}}\ \,q_1\in[\langle\Psi_1^{\parallel}|\Psi_1^{\parallel}\rangle,\ 1],
\end{equation}
where $|\Psi_1^{\parallel}\rangle$ is the component of $|\Psi_1\rangle$ which lies in the subspaces spanned by $\{|\Psi_{1'}\rangle$...$|\Psi_{N'}\rangle\}$.
The constraint (\ref{inner product}) for the quantum filtering is acquired based on the semidefinite property of the Gram matrix given by the vectors $\{|\Phi_1\rangle,|\Phi_{1'}\rangle,...,|\Phi_{i'}\rangle\}$ \cite{Bergou2005PRA}.
Set $S=(s_{11'},s_{12'},...,s_{1N'})$. We have the optimal solution,
\begin{subequations}\label{Succeeding of mixed state}
\begin{align}
\mathrm{(i):\ }&Q_{\min}\!\!=\!\!2\sqrt{P_1P_2\!\!\sum\limits_{i=1}^N\beta_is_{1i'}^2\!},\!\! &{\rm{when}}\ S\in\Lambda_{(i)};   \\
\mathrm{(ii):\ }&Q_{\min}\!\!=\!\!P_1\!\langle\Psi_1^{\parallel}\!|\!\Psi_1^{\parallel}\rangle\!\!+\!\!\frac{P_2\!(\!\sum\limits_{i=1}^N\beta_i s_{1i'}^2)\!\!}{\!\langle\Psi_1^{\parallel}|\Psi_1^{\parallel}\rangle\!},\!\! &{\rm{when}}\ S\in\Lambda_{(ii)};  \\
\mathrm{(iii):\ }&Q_{\min}\!=\!P_1\!+\!P_2(\sum\limits_{i=1}^N\beta_is_{1i'}^2),\!\! &{\rm{when}}\ S\in\Lambda_{(iii)},
\end{align}
\end{subequations}
where
\begin{subequations}\label{mixed state}
\begin{align}
&\Lambda_{(i)}=\{S:\ \langle\Psi_1^{\parallel}|\Psi_1^{\parallel}\rangle\leq q_1^*\leq1\},  \\
&\Lambda_{(ii)}=\{S:\ 0<q_1^*<\langle\Psi_1^{\parallel}|\Psi_1^{\parallel}\rangle\},  \\
&\Lambda_{(iii)}=\{S:\ q_1^*>1\},
\end{align}
\end{subequations}
and
\begin{equation}\label{parameter of filtering}
q_1^*=\sqrt{\frac{P_2}{P_1}\sum\limits_{i=1}^N\beta_is_{1i'}^2}.
\end{equation}
Corresponding to this optimal solution, for case (i) and (ii), both $|\Psi_1\rangle$ and $|\Psi_{i'}\rangle$ ($i=1,2,...,N$) are identified; for case (iii), the state $|\Psi_1\rangle$ is required to be neglected.

Since the fidelity has been found to be closely correlated with the state discrimination problems \cite{Josza1994}, Terry \emph{et al} \cite{Terry2003PRA} gives a lower bound of the optimal failure probability for the quantum filtering scheme,
\begin{equation}\label{fidelity lower bound}
Q_{\rm{min}}=2\sqrt{P_1P_2}F(\rho_1,\rho_2),
\end{equation}
where $F(\rho_1,\rho_2)$ is the fidelity between $\rho_1$ and $\rho_2$ given by \cite{Terry2003PRA},
\begin{eqnarray}\label{coherent state111}
F(\rho_1,\rho_2)&=&F(|\Psi_1\rangle\langle\Psi_1|,\rho_2) \nonumber\\
&=&\sqrt{\langle\Psi_1|\rho_2|\Psi_1\rangle}=\sqrt{\sum\limits_{i=1}^N\beta_is_{1i'}^2}.
\end{eqnarray}
 One can see that this lower bound is saturated for case (i) in Eq. (\ref{Succeeding of mixed state}a).

To see the essential difference between quantum superposition and classical mixture, and the role played by quantum coherence in high-dimensional mixed state discrimination, we replace the classical probability $\beta_i$ with quantum probability amplitudes \cite{JHZhang2020PRA}.
Thus, the vectors $|\Psi_i'\rangle$ ($i=1,...,N$) in the decomposition of $\rho_2$ are then superposed into a pure state,
\begin{equation}\label{coherent state110}
|\Psi_2\rangle=\sum\limits_{i=1}^N\sqrt{\beta_i}e^{i\theta_i}|\Psi_{i'}\rangle,
\end{equation}
where $0\leq\theta_i\leq2\pi$.

After the first results on characterization and quantification of coherence \cite{Aberg2014PRL,Levi2014NJP}, Baumgratz et al. \cite{Baumgratz2014PRL} put forward the resource-theoretic framework of coherence and formulated a set of axioms or preconditions for a measure of
coherence. As the bona fide measures for coherence, the $l_1$ norm of coherence is defined by
\begin{equation}
C_{l_1}(\rho)=\sum\limits_{i\neq j}|\rho_{ij}|,
\end{equation}
and the relative entropy of coherence is given by
\begin{equation}\label{relative entropy coherence}
C_{\rm{rel}}(\rho)=S(\rho_{\rm{diag}})-S(\rho),
\end{equation}
where $\rho=\sum_{ij}\rho_{ij}|i\rangle\langle j|$ is the density matrix and $\rho_{\rm{diag}}=\sum_{ii}\rho_{ii}|i\rangle\langle i|$
is the diagonal part of $\rho$. Both $l_1$ norm and relative entropy coherence measures are bases dependent. Below we consider the coherence under the fixed orthogonal basis $\{|\Psi_{i'}\rangle\}$ given in $\rho_2$. Thus, the $l_1$ norm coherence of $|\Psi_2\rangle$ is given by
\begin{equation}
C_{l_1}(|\Psi_2\rangle)=\sum\limits_{i\neq j}|\rho_{ij}|=
2\sum\limits_{i>j}^N\sqrt{\beta_i\beta_j}.
\end{equation}

\begin{figure}
%[!htbp]
\centering
\includegraphics[width=8cm]{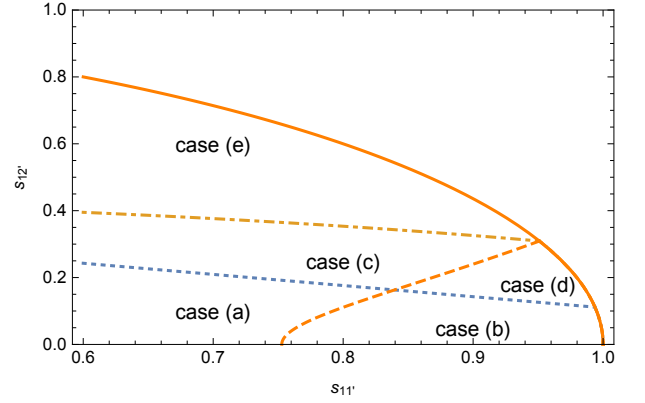} \\
\caption{Set $P_1=0.15$ and $\beta_1=0.1$. For $N=2$, we have five regions corresponding to cases (a)-(e), respectively, with respect to different values of $s_{11'}$ and $s_{12'}$. The dashed line for $q_1^*=s_{11'}^2+s_{12'}^2$, dot-dashed line for $q_1^*=1$, dotted line for $s^*=\sqrt{\frac{P_1}{P_2}}$ and orange solid line for $s_{11'}^2+s_{12'}^2=1$.} \label{fig1}
\end{figure}

The failure probability corresponding to the optimal discrimination between $|\Psi_1\rangle$ and $|\Psi_2\rangle$ is
\begin{subequations}\label{Succeeding of pure state}
\begin{align}
\mathrm{(i'):\, }&Q'_{\min}=2\sqrt{P_1P_2}|s^*|,\!\! &{\rm{when}}\ S\in\Lambda_{(i')},   \\
\mathrm{(ii'):\, }&Q'_{\min}=P_1+P_2|s^*|^2,\!\! &{\rm{when}}\ S\in\Lambda_{(ii')},
\end{align}
\end{subequations}
where
\begin{eqnarray}\label{coherent state112}
s^*&=&\langle\Psi_1|\Psi_2\rangle=\sum\limits_{i=1}^N\sqrt{\beta_i}s_{1i'}e^{i\theta_i},\nonumber\\
\Lambda_{(i')}&=&\{S:0\leq |s^*|\leq\sqrt{\frac{P_1}{P_2}}\},\nonumber\\
\Lambda_{(ii')}&=&\{S:\sqrt{\frac{P_1}{P_2}}<|s^*|\leq1\}.
\end{eqnarray}
Focusing on the difference between the result of classical mixture and quantum superposition, we consider
$\Delta Q=Q_{\min}-Q'_{\min}$ with respect to the following five cases, as the case $S\in\Lambda_{(i')}\cap\Lambda_{(iii)}$ corresponds to an empty set
according to (\ref{mixed state}) and (\ref{coherent state112}),
\begin{eqnarray}\label{five cases}
{\rm{case \ (a):}}\ S&\in&\Lambda_{(i')}\cap\Lambda_{(i)};~  {\rm{case\ (b):}}\  S\in\Lambda_{(i')}\cap\Lambda_{(ii)}; \nonumber\\
{\rm{case \ (c):}}\ S&\in&\Lambda_{(ii')}\cap\Lambda_{(i)};\,  {\rm{case \ (d):}}\  S\in\Lambda_{(ii')}\cap\Lambda_{(ii)}; \nonumber\\
{\rm{case \ (e):}}\ S&\in&\Lambda_{(ii')}\cap\Lambda_{(iii)},
\end{eqnarray}
see Fig. \ref{fig1} for $N=2$, $P_1=0.15$ and $\beta_1=0.1$.

To find out the role played by quantum supposition in our state discrimination, we consider the difference of the optimal success probability
between the pure-pure and pure-mixed state discrimination. We have the following theorem.

\emph{[Theorem 1].} The minimum failure probability $Q'_{\rm{min}}$ of pure-pure state discrimination is upper bounded by the one of quantum state filtering $Q_{\rm{min}}$, namely, $\Delta Q=Q_{\rm{min}}-Q_{\rm{min}}'\geq0$, if the following equal-fidelity condition holds, $F(\rho_1,\rho_2)=F(|\Psi_1\rangle,|\Psi_2\rangle)$.

%The failure probability ($Q'$) of coherent pure state discrimination is superior to the one of quantum state filtering ($Q$) except case (b). For $s_{1i'}=s_{1j'}=s_0$, the difference $\Delta Q=Q_{\min}-Q'_{\min}$ between the optimal probability (its square) is  proportional to $C_{l_1}(|\Psi_1\rangle)$ for case (e) (case (a)); its upper bounds is proportional to $C_{l_1}(|\Psi_1\rangle)$ for cases (c) and (d). Nevertheless, for case (b), there exists some cases where $\Delta Q\geq0$.

\emph{[Proof].} Since the fidelity between the two pure states $|\Psi_1\rangle$ and $|\Psi_2\rangle$ is given by
\begin{equation}\label{pure state fidelity}
F(|\Psi_1\rangle,|\Psi_2\rangle)=|s^*|,
\end{equation}
combining with Eqs. (\ref{Succeeding of mixed state}), (\ref{parameter of filtering}), (\ref{coherent state111}), (\ref{Succeeding of pure state}) and
(\ref{coherent state112}), we have the following results corresponding to the five different cases listed in (\ref{five cases}).

With respect to the case (a), we have
\begin{eqnarray}\label{case (a)}
(\Delta Q)^2&=&Q_{\rm{min}}^2-Q_{\rm{min}}'^2 \nonumber\\
   &=&4P_1P_2\sum\limits_{i=1}^N\beta_is_{1i'}^2-4P_1P_2|s^*|^2\nonumber\\
   &=&4P_1P_2[F^2(\rho_1,\rho_2)\!-\!F^2(|\Psi_1\rangle,|\Psi_2\rangle)]\!=\!0.
\end{eqnarray}

%For $s_{1i'}=s_{1j'}=s_0$, we have
%\begin{equation}
%\Delta Q^2=-4P_1P_2s_{0}^2\sum\limits_{i\neq j}^N\sqrt{\beta_i\beta_j}=-4P_1P_2s_0^2C_{l_1}(|\Psi_2\rangle).
%\end{equation}
For the case (b), we have
\begin{eqnarray}\label{negative}
\Delta Q&=&P_1\langle\Psi_1^{\parallel}|\Psi_1^{\parallel}\rangle+P_2\frac{\sum\limits_{i=1}^N\beta_is_{1i'}^2}{\langle\Psi_1^{\parallel}|\Psi_1^{\parallel}\rangle}\!-\!2\sqrt{P_1P_2}|s^*|\nonumber\\
&=&P_1\!\langle\Psi_1^{\parallel}|\Psi_1^{\parallel}\!\rangle\!+\!P_2\frac{F^2(\rho_1,\rho_2)}{\langle\Psi_1^{\parallel}|\Psi_1^{\parallel}\rangle}\!-\!2\!\sqrt{P_1P_2}\!F(\!|\Psi_1\rangle\!,\!|\Psi_2\rangle\!)\nonumber\\
&=&[\sqrt{P_1\langle\Psi_1^{\parallel}|\Psi_1^{\parallel}\rangle}-\sqrt{\frac{P_2}{\langle\Psi_1^{\parallel}|\Psi_1^{\parallel}\rangle}}F(\rho_1,\rho_2)]^2\geq0.
\end{eqnarray}

Corresponding to the case (c), we obtain that
\begin{eqnarray}\label{case (c)}
 S&\in&\Lambda_{(ii')}\cap\Lambda_{(i)}\nonumber\\
 &=&\{S:|s^*|>\sqrt{\frac{P_1}{P_2}}\}\cap\{S:\langle\Psi_1^{\parallel}|\Psi_1^{\parallel}\rangle\leq q_1^*\leq1\}\nonumber\\
 &=&\{S:F(|\Psi_1\rangle,|\Psi_2\rangle)>\sqrt{\frac{P_1}{P_2}}\}\nonumber\\
 &\cap&\{S:\sqrt{\frac{P_1}{P_2}}\langle\!\Psi_1^{\parallel}|\Psi_1^{\parallel}\rangle\!\leq\!F(\rho_1,\rho_2)\!\leq\!\sqrt{\frac{P_1}{P_2}}\},
\end{eqnarray}
which is just a empty set under the equal-fidelity condition.

 %\{s_{1t'}:\sqrt{\beta_t}s_{1t'}>\sqrt{\frac{P_1}{P_2}}\}\nonumber\\
 %&\cap&\{s_{1t'}:\sqrt{\beta_t}s_{1t'}<\sqrt{\frac{P_1}{P_2}}\}

%Case (c):
%\begin{eqnarray}
%\Delta %Q&=&P_1\langle\Psi_1^{\parallel}|\Psi_1^{\parallel}\rangle\!+\!P_2\frac{\sum\limits_{i=1}^N\beta_is_{1i'}^2}{\langle\Psi_1^{\parallel}|\Psi_1^{\parallel}\rangle}-(P_1\!+\!P_2s^{*2}).
%end{eqnarray}
%According to $\langle\Psi_1^{\parallel}|\Psi_1^{\parallel}\rangle\leq1$ and (\ref{mixed state}(a)), we have
%\begin{eqnarray}
%\Delta Q&\leq&P_1+P_2\sum\limits_{i=1}^N\beta_is_{1i'}^2-[P_1+P_2(\sum\limits_{i=1}^N\sqrt{\beta_i}s_{1i'})^2]\nonumber\\
%&=&-P_2\sum\limits_{i\neq j}^N\sqrt{\beta_i\beta_j}s_{1i'}s_{1j'}\leq0.
%\end{eqnarray}

For the case (d) we have
\begin{eqnarray}\label{case (d)}
 S&\in&\Lambda_{(ii')}\cap\Lambda_{(ii)}\nonumber\\
 &=&\{S:\sqrt{\frac{P_1}{P_2}}<|s^*|\leq1\}\cap\{S:0<q_1^*<\langle\Psi_1^{\parallel}|\Psi_1^{\parallel}\rangle\}\nonumber\\
 &=&\{S:F(|\Psi_1\rangle,|\Psi_2\rangle)>\sqrt{\frac{P_1}{P_2}}\}\nonumber\\
 &\cap&\{S:0<F(\rho_1,\rho_2)<\sqrt{\frac{P_1}{P_2}}\langle\Psi_1^{\parallel}|\Psi_1^{\parallel}\rangle\},
\end{eqnarray}
which is again a empty set under the equal-fidelity condition.

%Case (d):
%\begin{eqnarray}
%\Delta Q&=&2\sqrt{P_1P_2\sum\limits_{i=1}^N\beta_is_{1i'}^2}-[P_1+P_2(\sum\limits_{i=1}^N\sqrt{\beta_i}s_{1i'})^2]\nonumber\\
%&\leq&P_1+P_2\sum\limits_{i=1}^N\beta_is_{1i'}^2-[P_1+P_2(\sum\limits_{i=1}^N\sqrt{\beta_i}s_{1i'})^2]\nonumber\\
%&=&-P_2\sum\limits_{i\neq j}^N\sqrt{\beta_i\beta_j}s_{1i'}s_{1j'}\leq0.
%\end{eqnarray}

With respect to the case (e), we get
\begin{eqnarray}\label{case (e)}
\Delta Q&=&P_1+P_2\sum\limits_{i=1}^N\beta_is_{1i'}^2-(P_1+P_2|s^*|^2)\nonumber\\
&=&P_2[F^2(\rho_1,\rho_2)-F^2(|\Psi_1\rangle,|\Psi_2\rangle)]=0.
\end{eqnarray}

From the above results, we have that $\Delta Q=Q_{\rm{min}}-Q_{\rm{min}}'\geq0$ under the equal-fidelity condition $F(\rho_1,\rho_2)=F(|\Psi_1\rangle,|\Psi_2\rangle)$.
\hfill$\Box$

From the proof of Theorem 1, we see that the superiority of pure-pure state scheme versus pure-mixed one may only possibly occur for case (b).
Concerning the equal-fidelity condition in Theorem 1, we have the following conclusion.

\emph{[Corollary].}
The equal-fidelity condition $F(|\Psi_1\rangle,|\Psi_2\rangle)=F(\rho_1,\rho_2)$ in the simulation of quantum filtering is satisfied if and only if
\begin{equation}\label{expression for equal-fidelity condition}
\sum\limits_{i>j}^N\sqrt{\beta_i\beta_j}s_{1i'}s_{1j'}\cos(\theta_i-\theta_j)=0.
\end{equation}

\begin{figure}
%[!htbp]
\centering
\includegraphics[width=8cm]{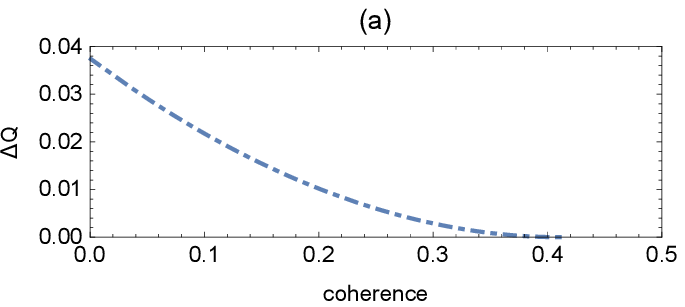} \\
\includegraphics[width=8cm]{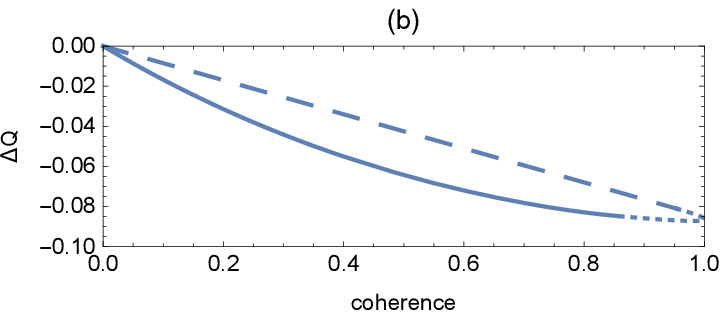} \\
\includegraphics[width=8cm]{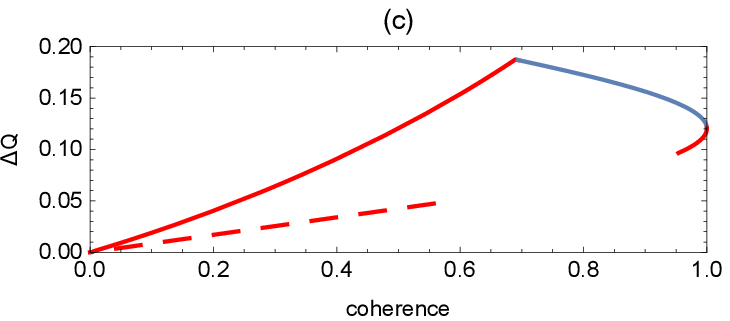} \\
 \caption{The difference $\Delta Q$ of the optimal success probabilities between the pure-mixed and pure-pure schemes as functions of the coherence encoded in the state $|\Psi_2\rangle$ for $N=2$, $P_1=0.15$ and $s_{12'}=0.5$. (a), (b) and (c) correspond to equal-fidelity case ($s_{11'}=0$), equal-phase case ($s_{11'}=0.2$, $\theta_1=\theta_2=0$) and unequal-phase case ($s_{11'}=0.2$, $\theta_1=\pi/2$, $\theta_2=-\pi/2$), respectively. Solid, dotted, dashed and dot-dashed lines correspond to the cases (a), (b), (d) and (e) in Fig. 1, respectively; while the case (c) does not match with here. Blue (red) lines correspond to the coherence detrimental (helpful) to state discrimination (the same for Figs. \ref{fig3}-\ref{fig7}).} \label{fig2}
\end{figure}

As for an illustration, consider $N=2$, $s_{1i'}\neq0$ and $s_{1j'}\neq0$. According to Eq. (\ref{expression for equal-fidelity condition}), we have
\begin{equation}\label{equal phase factor}
\cos(\theta_1-\theta_2)=0.
\end{equation}
Then, from Eqs. (\ref{mixed state}), (\ref{five cases}) and (\ref{equal phase factor}), the case (b) is also rejected.
Namely, $\Delta Q>0$ is impossible in this situation.
As for another example, let us consider the following case,
\begin{equation} \label{equal-fidelity example 1}
s_{1i'}=\delta_{it}s_{1t},
\end{equation}
which satisfies the equal-fidelity relation (\ref{expression for equal-fidelity condition}) obviously. We have
$$
\Delta Q=(\sqrt{P_1}s_{1t'}-\sqrt{P_2\beta_t})^2.
$$
Fig. \ref{fig2} (a) shows the relations between $\Delta Q$ and the coherence in this case.

Instead of the equal-fidelity condition, if we set the all phase factors in Eq. (\ref{coherent state110}) to equal to each other, $\theta_i=\theta_j$ ($i\neq j$, and $i,j=0, 1, 2, ..., N$), then we have the following theorem.

\emph{[Theorem 2].}  If $\theta_i=\theta_j$, the pure-pure state discrimination scheme is inferior to quantum state filtering, i.e., $\Delta Q=Q_{\min}-Q'_{\min}\leq0$ for all the cases except for the case (b). When $s_{1i'}=s_{1j'}=s_0$, $\Delta Q$ ($\Delta Q^2$) is proportional to $C_{l_1}(|\Psi_1\rangle)$ for the case (e) (case (a)), and the upper bound of $\Delta Q$ is proportional to $C_{l_1}(|\Psi_1\rangle)$ for cases (c) and (d).

%不用证明了。
\emph{[Proof].} For cases (a), (b) and (d), the expressions of $\Delta Q$ are the same as the ones in Eq. (\ref{case (a)}), (\ref{negative}) and (\ref{case (e)}). For the case (a), we have
\begin{eqnarray}
\Delta Q^2&=&Q_{\rm{min}}^2-Q_{\rm{min}}'^2 \nonumber\\
&=&-4P_1P_2\sum\limits_{i\neq j}\sqrt{\beta_i\beta_j}s_{1i'}s_{1j'}e^{i(\theta_i-\theta_j)} \nonumber\\
&=&-8P_1P_2\sum\limits_{i>j}\sqrt{\beta_i\beta_j}s_{1i'}s_{1j'}.\nonumber
\end{eqnarray}
For $s_{1i'}=s_{1j'}=s_0$, we have
\begin{eqnarray}
\Delta Q^2&=&-8P_1P_2s_{0}^2\sum\limits_{i>j}^N\sqrt{\beta_i\beta_j}\nonumber\\
&=&-4P_1P_2s_0^2C_{l_1}(|\Psi_2\rangle)<0.
\end{eqnarray}

Similarly, for the case (c), we have
$$
\Delta Q=P_1\langle\Psi_1^{\parallel}|\Psi_1^{\parallel}\rangle
+P_2\frac{\sum\limits_{i=1}^N\beta_is_{1i'}^2}{\langle\Psi_1^{\parallel}|
\Psi_1^{\parallel}\rangle}-(P_1\!+\!P_2|s^*|^2).
$$
According to that
$\langle\Psi_1^{\parallel}|\Psi_1^{\parallel}\rangle\leq1$, %and (\ref{coherent state112}),
we have
\begin{eqnarray}
\Delta Q&\leq&P_1+P_2\sum\limits_{i=1}^N\beta_is_{1i'}^2-(P_1+P_2|\sum\limits_{i=1}^N\sqrt{\beta_i}s_{1i'}e^{i\theta_i}|^2)\nonumber\\
&=&-2P_2\sum\limits_{i>j}^N\sqrt{\beta_i\beta_j}s_{1i'}s_{1j'}\cos(\theta_i-\theta_j)\nonumber\\
&=&-P_2s_0^2C_{l_1}(|\Psi_2\rangle)<0.\nonumber
\end{eqnarray}

For the case (d), we get
\begin{eqnarray}
\Delta Q&=&2\sqrt{P_1P_2\sum\limits_{i=1}^N\beta_is_{1i'}^2}-P_1-P_2|\sum\limits_{i=1}^N\sqrt{\beta_i}s_{1i'}e^{i\theta_i}|^2\nonumber\\
&\leq&P_1+P_2\sum\limits_{i=1}^N\beta_is_{1i'}^2-P_1-P_2|\sum\limits_{i=1}^N\sqrt{\beta_i}s_{1i'}e^{i\theta_i}|^2\nonumber\\
&=&-2P_2\sum\limits_{i>j}^N\sqrt{\beta_i\beta_j}s_{1i'}s_{1j'}\cos(\theta_i-\theta_j)\nonumber\\
&=&-P_2s_0^2C_{l_1}(|\Psi_2\rangle)<0.\nonumber
\end{eqnarray}

For the case (e), we obtain
\begin{eqnarray}
\Delta Q&=&P_1+P_2\sum\limits_{i=1}^N\beta_is_{1i'}^2-[P_1+P_2|\sum\limits_{i=1}^N\sqrt{\beta_i}s_{1i'}e^{i\theta_i}|^2]\nonumber\\
&=&-2P_2\sum\limits_{i>j}^N\sqrt{\beta_i\beta_j}s_{1i'}s_{1j'}\cos(\theta_i-\theta_j)\nonumber\\
&=&-P_2s_0^2C_{l_1}(|\Psi_2\rangle)<0.\nonumber
\end{eqnarray}
%引理1中可加入有关相位角的讨论。
%保真度、相干叠加相因子、一一对应内积条件三者"契合“关系以及相干是否有益于态区分讨论
%分别取N=2，N=3的情况进行讨论。
%此处还可以通过调节相因子实现几率反超现象，但本质上是通过改变保真度导致相应态区分成功几率的改变。
\hfill$\Box$

%(此处考虑加一个命题1：相因子等于零。叠加态保真度一定大于等于纯-混态保真度，反之则结论相反）

The inequality $\Delta Q<0$ does not always hold under the condition in Theorem 2 for
the case (b) can be seen from Theorem 1, where it is indicated that $\Delta Q\geq0$
under the equal-fidelity condition.
To illustrate the role played by the quantum coherence in our procedure, we show the difference $\Delta Q$ as a function of coherence of $|\Psi_2\rangle$ in Fig. \ref{fig2}. One can see that for the equal-phase cases shown in Figs. \ref{fig2} (a) and (b), the quantum coherence is not a critical recourse but detrimental to the unambiguous state discrimination even for the cases where pure-pure state scheme is superior to pure-mixed one as guaranteed by the equal-fidelity condition. This result is different from the one in \cite{Kim2018arXiv} where the coherence generated in the auxiliary system is positively correlated with the optimal success probability of state discrimination.

Nevertheless, when the phase factors $\theta_i$ turn to be unequal (shown in Fig. \ref{fig2} (c)), the following two conclusions may be drawn:
(i) the optimal success probability of pure-mixed scheme may be surpassed by pure-pure state one on the contrary; (ii) some of the coherence encoded in the pure state is not detrimental but helpful to state discrimination (shown in Fig. \ref{fig2} (c)). Namely, one can acquire helpful coherence via adjusting the phase factors in the supposed state $|\Psi_2\rangle$. By a straightforward calculation, it is easily known that this superiority of the pure-pure state discrimination scheme versus the pure-mixed one, as shown in Fig. \ref{fig2} (c), can be attributed to the fact that $F(\rho_1,\rho_2)\geq F(|\Psi_1\rangle, |\Psi_2\rangle)$ according to Eqs. (\ref{coherent state111}) and (\ref{pure state fidelity}). The reverse is also true for the results in Fig. \ref{fig2} (b). Since the state with a lower fidelity is easier to be discriminated, the superiority of pure-pure state discrimination versus pure-mixed one occurs without surprising. Then, we will focus on the superiority of pure-pure scheme versus pure-mixed one under equal-fidelity cases in the next section.

\section{Discrimination of two rank-$N$ mixed states}\label{Simulation1}

We have studied the quantum filtering problem for the discrimination of two special rank-$N$ mixed states. The results for quantum filtering indicates that pure-pure state discrimination scheme with a same fidelity as the pure-mixed one tends to be more possible to succeed. This prompts us to investigate the discrimination of two rank-$N$ mixed states under the equal-fidelity condition. The two mixed states are of the following form,
\begin{equation}\label{rank-N mixed state}
\rho_1=\sum\limits_{i=1}^N\alpha_i|\Psi_i\rangle\langle\Psi_i|,
~~~\rho_2=\sum\limits_{i=1}^N\beta_i|\Psi_{i'}\rangle\langle\Psi_{i'}|,
\end{equation}
where $\{|\Psi_i\rangle\}$ and $\{|\Psi_{i'}\rangle\}$ are orthonormal bases which satisfy
\begin{equation}\label{delta equation}
\langle\Psi_i|\Psi_j\rangle\!=\!\delta_{ij},~\langle\Psi_{i'}
|\Psi_{j'}\rangle\!=\!\delta_{ij},~\langle\Psi_i|\Psi_{j'}\rangle\!=\!s_{ii'}\delta_{ij}
\end{equation}
with $i,j=1,...,N$ and $s_{ii'}=\langle\Psi_i|\Psi_{i'}\rangle>0$. The state $\rho_i$ occurs with \emph{a priori} probability $P_i$ ($i=1,2$, $P_1\alpha_i\leq P_2\beta_i$).

The relation (\ref{delta equation}) means that the vectors composing $\rho_1$ are one to one overlapped with the ones of $\rho_2$.
The relation (\ref{delta equation}) is satisfied for the following example:
\begin{equation}\label{Example of one to one overlapping case}
|\Psi_i\rangle\!=\!|2i\!-\!2\rangle,~~~|\Psi_{i'}\rangle\!=\!s_{ii'}|2i\!-\!2\rangle+\sqrt{1\!-\!|s_{ii'}|^2}|2i\!-\!1\rangle,
\end{equation}
where $\{|2i-1\rangle\}$ and $\{|2i\rangle\}$ ($i=1,...,N$) are orthonormal bases in a $2N$ dimensional Hilbert space.

If we compare the results with this discrimination of a pair of pure states,
\begin{equation}\label{N dimensional pure state}
|\Phi_1\rangle=\sum\limits_{i=1}^N\sqrt{\alpha_i}|\Psi_i\rangle,~~~
|\Phi_2\rangle=\sum\limits_{i=1}^N\sqrt{\beta_i}|\Psi_{i'}\rangle,
\end{equation}
occurring with \emph{a priori} probability $P_1$ and $P_2$, respectively,
the relation (\ref{delta equation}) guarantees the equal-fidelity condition $$F(\rho_1,\rho_2)=F(|\Phi_1\rangle,|\Phi_2\rangle)
=\sum\limits_{i=1}^N\sqrt{\alpha_i\beta_i}\langle\Psi_i|\Psi_{i'}\rangle.$$

The conditions in Eq. (\ref{delta equation}) also ensure that the discrimination of $\rho_1$ and $\rho_2$ can be carried out in $N$ independent subspaces through optimal POVM operators which can be written as a direct sum of $N$ corresponding parts, just like the results for the discrimination of rank-two mixed state in \cite{Namkung2017PRA,JHZhang2020PRA}. Then, concerning the optimal discrimination of $\rho_1$ and $\rho_2$, we have the following remark.

\emph{[Remark].} The successful probability of discrimination between the two rank-$N$ states $\rho_1$ and $\rho_2$ in Eq. (\ref{rank-N mixed state}) satisfying the relation (\ref{delta equation}) is equivalent to a weighed average of the one for the discrimination between the $i$th pair of eigenvectors $\{|\Psi_{i}\rangle,|\Psi_{i'}\rangle\}$ ($i=1,...,N$).

Thus, if  $0<s_{ii'}\leq\sqrt{\frac{P_1\alpha_i}{P_2\beta_i}}$ ($1\leq i\leq m$), $\sqrt{\frac{P_1\alpha_i}{P_2\beta_i}}<s_{ii'}\leq1$ ($m+1\leq i\leq N$), where $m$ is an integer satisfies $1<m<N$, the minimum failure probability for discriminating $\rho_1$ from $\rho_2$ is given by
$$
Q_{\rm{min}}=\sum\limits_{i=1}^m2\sqrt{P_1P_2\alpha_i\beta_i}s_{ii'}\!
+\!\sum\limits_{i=m+1}^N(P_1\alpha_i\!+\!P_2\beta_is_{ii'}^2).
$$
Here, the vectors \{$|\Psi_i\rangle,|\Psi_{i'}\rangle$\} ($1\leq i\leq m$) are all identified while $|\Psi_i\rangle$ ($m<i\leq N$) are neglected in the optimal solution for the discrimination between $\rho_1$ and $\rho_2$.

For the discrimination of pure-pure states, optimal failure probability is of the same form as Eq. (\ref{Succeeding of pure state}). Set
\begin{equation}
s^*=\sum\limits_{i=1}^N\sqrt{\alpha_i\beta_i}s_{ii'}.
\end{equation}
 We have the following theorem as an extension of the work for the discrimination of rank-two mixed states in \cite{JHZhang2020PRA}.

\emph{[Theorem 3].} For the discrimination of two rank-$N$ mixed states in Eq. (\ref{rank-N mixed state}), the minimum failure probability $Q_{\rm{min}}'$ corresponding to the optimal discrimination of the pure states in Eq. (\ref{N dimensional pure state}) is upper bounded by $Q_{\rm{min}}$ for the mixed states.

\emph{[Proof].}  Corresponding to different values of $s_{ii'}$, we have the following four cases.

Case (i): $0<s_{ii'}\leq\sqrt{\frac{P_1\alpha_i}{P_2\beta_i}}$, which implies that
$s^*\leq\sqrt{\frac{P_1}{P_2}}$ ($i=1,2,...,N$). Here, all vectors included in $\rho_1$ and $\rho_2$ are identified. We have
\begin{eqnarray}\nonumber
Q_{\rm{min}}&=&\sum\limits_{i=1}^N\sqrt{P_1P_2}\sqrt{\alpha_i\beta_i}s_{ii'}
=\sqrt{P_1P_2}\sum\limits_{i=1}^N\sqrt{\alpha_i\beta_i}s_{ii'}\nonumber\\
&=&\sqrt{P_1P_2}s^*=Q'_{\rm{min}}.\nonumber
\end{eqnarray}

Case (ii):  $\sqrt{\frac{P_1\alpha_i}{P_2\beta_i}}<s_{ii'}<1$, which gives rise to $s^*>\sqrt{\frac{P_1\beta_i}{P_2\alpha_i}}$. All of the vectors included in $\rho_1$ are neglected in the optimal solution for discrimination of $\rho_1$ and $\rho_2$. According to the Cauchy-Schwarz inequality, the optimal failure probability for succeedingly discriminating $|\Psi_1\rangle$ from $|\Psi_2\rangle$ satisfies
\begin{eqnarray}
Q'_{\rm{min}}&=&P_1+P_2(\sum\limits_{i=1}^N\sqrt{\alpha_i\beta_i}s_{ii'})^2\nonumber\\
&\leq&P_1+P_2(\sum\limits_{j=1}^N\alpha_i)(\sum\limits_{i=1}^N\beta_is_{ii'}^2)\nonumber\\
&=&P_1+P_2(\sum\limits_{i=1}^N\beta_is_{ii'}^2)=Q_{\rm{min}}.
\end{eqnarray}
This upper bound is saturated when
$$\frac{\alpha_1}{\beta_1s^2_{11'}}=\frac{\alpha_2}{\beta_2s^2_{22'}}
=...=\frac{\alpha_N}{\beta_Ns^2_{NN'}}.$$

Case (iii): $0<s_{ii'}\leq\sqrt{\frac{P_1\alpha_i}{P_2\beta_i}}$ ($1\leq i\leq m$), $\sqrt{\frac{P_1\alpha_i}{P_2\beta_i}}<s_{ii'}\leq1$ ($m+1\leq i\leq N$) and $s^*\leq\sqrt{\frac{P_1}{P_2}}$, where $1<m<N$.

The difference $\Delta Q$ between the two schemes is given by
\begin{eqnarray}
\Delta Q&=&Q_{\rm{min}}-Q'_{\rm{min}}\nonumber\\
&=&\sum\limits_{i=1}^m2\sqrt{P_1P_2\alpha_i\beta_i}s_{ii'}\!+\!\sum\limits_{i=m+1}^N(P_1\alpha_i\!+\!P_2\beta_is_{ii'}^2)\nonumber\\
& &\!-\!2\sqrt{P_1P_2}\sum\limits_{i=1}^N\sqrt{\alpha_i\beta_i}s_{ii'}\nonumber\\
&=&\sum_{i=m+1}^N(\sqrt{P_1\alpha_i}-\sqrt{P_2\beta_i}s_{ii'})^2>0.
\end{eqnarray}

Case (iv): $0<s_{ii'}\leq\sqrt{\frac{P_1\alpha_i}{P_2\beta_i}}$ ($1\leq i\leq m$), $\sqrt{\frac{P_1\alpha_i}{P_2\beta_i}}<s_{ii'}\leq1$ ($m+1\leq i\leq N$) and $s^*>\sqrt{\frac{P_1}{P_2}}$. We have
\begin{eqnarray}
\Delta Q&=&\sum\limits_{i=1}^m2\sqrt{P_1P_2\alpha_i\beta_i}s_{ii'}\!+\!\sum\limits_{i=m+1}^N(P_1\alpha_i\!+\!P_2\beta_is_{ii'}^2)\!\nonumber\\
& &\!-\!P_1\!-\!P_2(\sum\limits_{i=1}^N\sqrt{\alpha_i\beta_i}s_{ii'})^2.
\end{eqnarray}

We prove $\Delta Q\geq0$ via mathematical induction. First, consider the case for $m=1$.
We have
\begin{eqnarray}\label{the first expression}
\Delta Q_1&=&\sqrt{P_1P_2\alpha_1\beta_1}s_{11'}\!+\!\sum\limits_{i=2}^N(P_1\alpha_i\!+\!P_2\beta_is_{ii'}^2)\!\nonumber\\
& &-P_1-P_2s^{*2}.
\end{eqnarray}

This expression is a quadric function of the variable $s_{11'}$ with a negative quadratic coefficient $-P_2\alpha_1\beta_1$.
Since $0<s_{11'}\leq\sqrt{\frac{P_1\alpha_1}{P_2\beta_1}}$, $\Delta Q$ achieves its minimum (lower limit) at the boundary points $s_{11'}\to0$ and $s_{11'}=\sqrt{\frac{P_1\alpha_1}{P_2\beta_1}}$.

Then, according to Eq. (\ref{the first expression}), $s^*>\sqrt{\frac{P_1}{P_2}}$, $s_{11'}\to0$,
$\sum\limits_{i=1}^N\beta_i=1$ and the Cauchy-Schwarz inequality,
we have
\begin{eqnarray}
\Delta Q_1&=&\sum\limits_{i=2}^NP_1\alpha_i\!+\!\frac{P_2\sum\limits_{i=2}^N\alpha_i\sum\limits_{i=2}^N\beta_is_{ii'}^2}{1-\alpha_1}\!-\!P_1\!-\!P_2s^{*2}\nonumber\\
&\geq&P_1(1\!-\!\alpha_1)\!+\!P_2(\frac{1}{1\!-\!\alpha_1}\!-\!1)\sum\limits_{i=2}^N\sqrt{\alpha_i\beta_i}s_{ii'}\!-\!P_1\nonumber\\
&>&-P_1\alpha_1+\frac{P_2\alpha_1}{1-\alpha_1}\sqrt{\frac{P_1}{P_2}}>-P_1\alpha_1+\frac{P_2\alpha_1}{1-\alpha_1}\frac{P_1}{P_2}\nonumber\\
&=&\frac{P_1\alpha_1^2}{1-\alpha_1}>0.
\end{eqnarray}

Corresponding to another boundary point $s_{11'}=\sqrt{\frac{P_1\alpha_1}{P_2\beta_1}}$, we have
\begin{eqnarray}
\Delta Q_1
&=&2\sqrt{\frac{P_1\alpha_1}{P_2\beta_1}}\sqrt{P_1\alpha_1P_2\beta_1}+\sum\limits_{i=2}^N(P_1\alpha_i+P_2\beta_is_{ii'}^2)\nonumber\\
& &-P_1-P_2(\sum\limits_{i=1}^N\sqrt{\alpha_i\beta_i}s_{ii'})^2\nonumber\\
&=&2P_1\alpha_1+P_1(1-\alpha_1)+\frac{P_2\sum\limits_{i=2}^N\alpha_i\sum\limits_{i=2}^N\beta_is_{ii'}^2}{1-\alpha_1}-P_1\nonumber\\
& &-P_2(\sqrt{\frac{P_1}{P_2}}\alpha_1+\sum\limits_{i=2}^N\sqrt{\alpha_i\beta_i}s_{ii'})^2\nonumber\\
&\geq&P_1\alpha_1+\frac{P_2}{1-\alpha_1}(\sum\limits_{i=2}^N\sqrt{\alpha_i\beta_i}s_{ii'})^2\nonumber\\
& &-P_2(\sqrt{\frac{P_1}{P_2}}\alpha_1+\sum\limits_{i=2}^N\sqrt{\alpha_i\beta_i}s_{ii'})^2\nonumber\\
&=&\frac{\alpha_1^2}{1-\alpha_1}[\sqrt{P_1}(1-\alpha_1)-\sqrt{P_2}\sum\limits_{i=2}^N\sqrt{\alpha_i\beta_i}s_{ii'}]^2\nonumber\\
&\geq&0.
\end{eqnarray}

As an induction hypothesis, we suppose that our conclusion holds for $m=k$, $1<k<N$,
\begin{eqnarray}\label{difference of Q}
\Delta Q_k&=&T\!\!+\!\!2\sqrt{P_1\alpha_kP_2\beta_k}s_{kk'}\!\!+\!\!P_1\alpha_{k\!+\!1}\!+\!P_2\beta_{k\!+\!1}s_{k\!+\!1,(k\!+\!1)'}^2\nonumber\\
& &+M-P_1-P_2(\sum_{i=1}^N\sqrt{\alpha_i\beta_i}s_{ii'})^2>0,
\end{eqnarray}
where
$$T\!=\!2\sqrt{P_1P_2}(\sum_{i=1}^{k-1}\sqrt{\alpha_i\beta_i}s_{ii'}),~~~
M\!=\!\sum_{i=k+2}^N(P_1\alpha_i\!+\!P_2\beta_is_{ii'}^2).$$
%$$T=2\sqrt{P_1P_2}(\sum_{i=1}^{k-1}\sqrt{\alpha_i\beta_i}s_{ii'})$$
%$$M=\sum_{i=k+2}^N(P_1\alpha_i\!+\!P_2\beta_is_{ii'}^2).$$
Then, for $m=k+1$, we have
\begin{eqnarray}
 \Delta Q_{k+1}&=&T\!\!+\!\!2\sqrt{P_1\alpha_kP_2\beta_k}s_{kk'}\!\!
 +\!\!2\sqrt{P_1\alpha_{k\!+\!1}\beta_{k\!+\!1}}s_{(k\!+\!1)',k\!+\!1}\nonumber\\
 & &+M-P_1-P_2(\sum_{i=1}^N\sqrt{\alpha_i\beta_i}s_{ii'})^2.\nonumber
\end{eqnarray}

Here, $\Delta Q_{k+1}$ can also be considered as a quadratic function of $x$ ($x=s_{(k+1)',k+1}$) with a minus coefficient of quadratic term. Thus, we also can acquire the minimum $\Delta Q_{\min}$ at the boundary points $x=0$ and $\sqrt{\frac{P_1\alpha_{k+1}}{P_2\beta_{k+1}}}$. For $x=0$, according to that $s^*>\sqrt{\frac{P_1}{P_2}}$, we have
\begin{eqnarray}
& &\Delta Q_{k+1}|_{x=0}-\Delta Q_{k}\bigg|_{x=\sqrt{\frac{P_1\alpha_{k+1}}{P_2\beta_{k+1}}}}\nonumber\\
&=&P_1\alpha_{k+1}^2+2\sqrt{P_1P_2}(\!\sum\limits_{i\neq k+1}\sqrt{\alpha_i\beta_i}s_{ii'})\alpha_{k+1}\!-2P_1\alpha_{k+1}\nonumber\\
&=&P_1\alpha_{k+1}^2+2\sqrt{P_1P_2}s^*\alpha_{k+1}\!-2P_1\alpha_{k+1}\nonumber\\
&>&P_1\alpha_{k+1}^2\geq0.
\end{eqnarray}
Hence, we have
$$\Delta Q_{k+1}\big|_{x=0}>\Delta Q_{k}\bigg|_{x=\sqrt{\frac{P_1\alpha_{k+1}}{P_2\beta_{k+1}}}}>0
$$
according to the relation (\ref{difference of Q}).
%Assume that both $Q$ and $\Delta Q$ are continuous functions of the parameter $x$.
For another boundary point $x=\sqrt{\frac{P_1\alpha_k}{P_2\beta_k}}$, we have
$$\Delta Q_{k\!+\!1}\bigg|_{x\!\to\!\sqrt{\frac{P_1\alpha_k}{P_2\beta_k}}}\!=\!\Delta Q_k\bigg|_{x\!=\!\sqrt{\frac{P_1\alpha_k}{P_2\beta_k}}}>0.$$
Therefore, $\Delta Q_{k+1}>0$. \hfill $\Box$

\begin{figure}
%[!htbp]
\centering
\includegraphics[width=8cm]{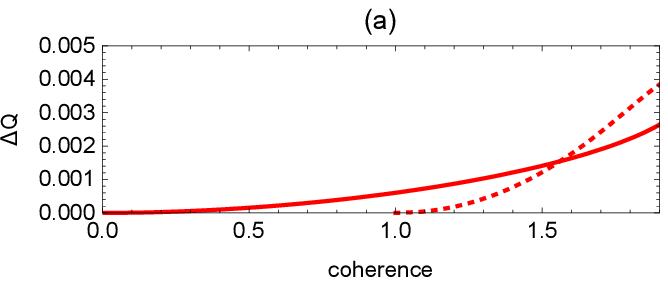} \\
\includegraphics[width=8cm]{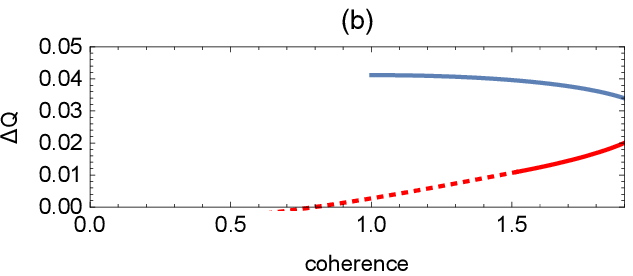} \\
\caption{The difference $\Delta Q$ of the minimum failure probability  between the two schemes as functions of the global coherence $C_{l_1}(|\Psi_i\rangle)$ corresponding to the cases for $N=3$, $P_1=0.15$ and $P_2=0.85$. (a) and (b) correspond to $s_{11'}=0.2$, $s_{22'}=0.5$, $s_{33'}=0.5$ and $s_{11'}=0.5$, $s_{22'}=0.2$, $s_{33'}=0.2$, respectively. The solid and dotted line corresponds to case (ii) and (iii) in \emph{Theorem 3}, respectively.} \label{fig3}
\end{figure}

Hence, it can be concluded that the discrimination of the pure superposed states is bound to be more possible to succeed than the mixed ones due to the equal-fidelity condition (\ref{delta equation}). Namely, the results of {\emph{{Theorem 1}}} %is italic necessary?
 in \cite{JHZhang2020PRA} can be generalized to this rank-$N$ system successfully.
Set $N=3$, $\beta_1=\alpha_1$ and $\alpha_i=\beta_i=\frac{1-\alpha_1}{2}$ ($i=2,3$). Here, different from filtering, coherence exists symmetrically in the two pure states $|\Phi_1\rangle$ and $|\Phi_2\rangle$.
Then, let us consider the difference $\Delta Q=Q_{\rm{min}}-Q'_{\rm{min}}$ as a function of the global coherence measured by the $l_1$ norm. It shows that there are much more non-vanishing and helpful coherence regions in which $\Delta Q>0$ than that for quantum filtering problems, see Figure \ref{fig3}.
The superiority of pure-pure scheme is inferior to the results of quantum filtering obviously.

%尝试将此图改为与无限维情形相对应的参数，
\section{Infinite-dimensional systems}\label{infinite dimensional}

Now, we aim to extend the above state discrimination problems to infinite-dimensional systems associated with two Gaussian states in quantum optics. First, we consider the following examples including the results for binomial states \cite{Stoler1985} as an intermediate transition from finite to infinite dimensional system problems.

\emph{[Example 1].  Equal-fidelity cases of quantum filtering.}
In this example, we discriminate a pure state $|\Psi_1\rangle$ from one of the following two states:\\
\noindent  (1) a rank-$N$ mixed state with the eigenvalues corresponding to the binomial distribution \cite{Stoler1985}, which is equivalent to the expression of the Poisson distribution for $N\to\infty$:
\begin{equation}\label{finite-dimensional mixed state}
\rho^0_2=\sum\limits_{i=0}^N |f_0(\alpha,i)|^2|\Psi_{i'}\rangle\langle\Psi_{i'}|,
\end{equation}
where
\begin{equation}\label{binomial distributio}
f_0(\alpha,i)=\sqrt{\frac{N!}{i!(N-i)!}(\frac{\alpha^2}{N})^i(1-\frac{\alpha^2}{N})^{N-i}};
\end{equation}
\noindent (2) a rank-$\infty$ mixed state:
\begin{equation}\label{infinite-dimensional mixed state}
\rho_{2}=\sum\limits_{i=0}^{\infty}|f(\alpha,i)|^2|\Psi_{i'}\rangle\langle\Psi_{i'}|,
\end{equation}
where the function $f(\alpha,i)$ corresponds to the photon number distribution of a given Gaussian state (the well-known coherent or squeezed vacuum state) which is notable in quantum optics. For the case associated with the well-known coherent state, we have
\begin{equation}\label{Poisson distribution}
f(\alpha,i)=e^{-|\alpha|^2/2}\alpha^i/\sqrt{i!},
\end{equation}
where $\alpha=|\alpha|e^{{\rm{i}}\theta}=re^{{\rm{i}}\theta}$;
for the one with respect to the squeezed vacuum state, one has
\begin{equation}\label{maxc}
f(\alpha,i)=\frac{[-e^{{\rm{i}}\theta}\tanh r(|\alpha|)]^i(2i!)^{1/2}}{\sqrt{\cosh r(|\alpha|)}i!2^i}
\end{equation}
with
\begin{equation}\label{equal photon number}
r(|\alpha|)=\ln(|\alpha|+\sqrt{|\alpha|^2+1}).
\end{equation}

As the binomial distribution of photon numbers is equivalent to Poisson distribution when $N\to\infty$, we obtain that $\lim\limits_{N\to\infty}|f_0(\alpha,i)|=\frac{|\alpha|^i}{\sqrt{i}!}e^{-|\alpha|^2/2}$. From the relation (\ref{equal photon number}), we see that $\sinh^2[r(|\alpha|)]=|\alpha|^2$, which guarantees the equivalence of average photon number between the generalized coherent and the squeezed vacuum states.

The relations (\ref{overlap}) and (\ref{equal-fidelity example 1}) are also satisfied for both $\rho^0_2$ and $\rho_{2}$. That is, the vector $|\Psi_1\rangle$ is only overlapped with the $t$th vector $|\Psi_{t'}\rangle$ in $\rho_2$ ($\rho_2^0$).  We discriminate the state $|\Psi_1\rangle$ from a superposed state $|\Psi^0_2\rangle$ and a generalized Gaussian state $|\Psi_2\rangle$ given by
\begin{eqnarray}\label{finite-dimensional pure state}
|\Psi_2^0\rangle &=&\sum\limits_{i=0}^Nf_0(\alpha,i)|\Psi_{i'}\rangle;\\
|\Psi_2\rangle &=&\sum\limits_{i=0}^{\infty} f(\alpha,i)|\Psi_{i'}\rangle.
\end{eqnarray}
One can easily obtain that $F(|\Psi_1\rangle\langle \Psi_1|,\rho_2^0)=F(|\Psi_1\rangle,|\Psi_2^0\rangle)$ and $F(|\Psi_1\rangle\langle \Psi_1|,\rho_2)=F(|\Psi_1\rangle,|\Psi_2\rangle)$, corresponding to the equal-fidelity condition (\ref{equal-fidelity example 1}).

\emph{[Example 2]. Equal-fidelity cases of two mixed state discrimination.} In this example, we consider the discrimination of the two pairs of states occurring with prior probabilities $P_1$ and $P_2$:

(1) rank-$N$ mixed states
\begin{eqnarray}\label{mixed finite state}
\rho_1^0&=&\sum\limits_{i=0}^N |f_0(\alpha,i)|^2|\Psi_{i}\rangle\langle\Psi_{i}|,\nonumber\\
\rho_2^0&=&\sum\limits_{i=0}^N |f_0(\alpha,i)|^2|\Psi_{i'}\rangle\langle\Psi_{i'}|;
\end{eqnarray}

(2) rank-$\infty$ mixed states
\begin{eqnarray}\label{mixed infinite state}
\rho_1&=&\sum\limits_{i=0}^{\infty}|f(\alpha,i)|^2|\Psi_i\rangle\langle\Psi_i|,\nonumber\\
\rho_2&=&\sum\limits_{i=0}^{\infty}|f(\alpha,i)|^2|\Psi_{i'}\rangle\langle \Psi_{i'}|,
\end{eqnarray}
where $|\Psi_i\rangle$ and $|\Psi_{i'}\rangle$ are orthonormal bases satisfying Eq. (\ref{delta equation}) for $0\leq i<\infty$.

%where $\{|i\rangle\}$ and {$\{|k_i\rangle\}$ are orthonormal basis which satisfies
%\begin{equation}\label{one to one overlap}
%\langle i|j\rangle=\delta_{i,j},~\langle {k_i}|{k_j}\rangle=\delta_{i,j},~\langle i|{k_j}\rangle=s_{i,{k_j}}\delta_{i,j}.
%\end{equation}

Then, we consider the discrimination of pure states with the two sets of bases superposed as follows:
\begin{eqnarray}\label{generalized coherent state}
|\Phi_1^0\rangle &\!=\!&\sum\limits_{i=0}^{N}\!f_0(\alpha,i)\!|\Psi_i\rangle,~~~|\Phi_2^0\rangle\!=\!\sum\limits_{i=0}^{N}\!f_0(\alpha,i)\!|\Psi_{i'}\rangle;\\
|\Phi_{1}\rangle &\!=\!&\sum\limits_{i=0}^{\infty}\!f(\alpha,i)\!|\Psi_i\rangle,~~~|\Phi_{2}\rangle\!=\!\sum\limits_{i=0}^{\infty} \!f(\alpha,i)\!|\Psi_{i'}\rangle,
\end{eqnarray}
where $|\Phi^0_j\rangle$ ($|\Phi_j\rangle$) ($j=1,2$) is a superposed binomial state (generalized Gaussian state) that satisfies $F(\rho_1^0,\rho_2^0)=F(|\Phi_1^0\rangle,|\Phi_2^0\rangle)$ ($F(\rho_1,\rho_2)=F(|\Phi_1\rangle,|\Phi_2\rangle)$) obviously.

%We can present an example of the physical realization of a quantum states satisfying Eq. (\ref{one to one overlap}).  In the framework of Janes-Cummings model \cite{Jaynes1963IEEE} which is notable model for characterizing the dynamics of interaction between two-level atoms and light field in quantum optics, one can
%derive the required state (\ref{generalized coherent state}) as follows:
%$$|i\rangle=|i\rangle_0\otimes|g\rangle,~~|k_i\rangle=s_{i,k_i}|i\rangle_0\otimes|g\rangle\!+\!\sqrt{1\!-\!|s_{i,k_i}|^2}|i\!+\!1\rangle_0\otimes|e\rangle,$$
%where $\{|i_0\rangle\}$ is Fock states of light field; $|e\rangle$ ($|g\rangle$) is the excited (ground) state of a two-level atom.

Concerning the role played by the coherence in above two examples, we choose the relative entropy coherence \cite{Baumgratz2014PRL,YRZhang2016PRA} defined by Eq. (\ref{relative entropy coherence}), since the $l_1$-norm coherence does't fulfill that the coherence is finite for $N\to\infty$. Calculating the global coherence of $|\Phi^0_j\rangle$ and $|\Phi_j\rangle$ ($j=1,2$) measured by the relative entropy, we have
\begin{eqnarray}
C_{\rm{rel}}&=&-\sum\limits_{i=0}^{N}|f_0(\alpha,i)|^2\log|f_0(\alpha,i)|^2;\nonumber\\
C_{\rm{rel}}&=&-\sum\limits_{i=0}^{\infty}|f(\alpha,i)|^2\log|f(\alpha,i)|^2.
\end{eqnarray}
%$$C_{\rm{rel}}=e^{-|\alpha|^2}\sum\limits_{n=0}^{\infty}\frac{|\alpha|^{2n}\log n!}{n!}-|\alpha|^2\log\frac{|\alpha|^2}{e}.$$

\begin{figure}
%[!htbp]
\centering
\includegraphics[width=8cm]{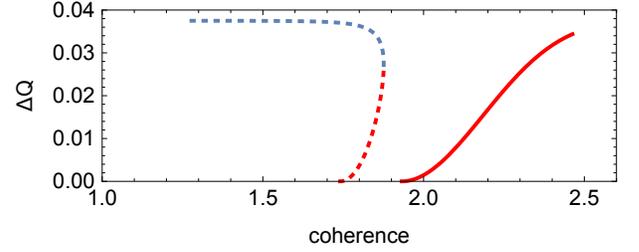} \\
 \caption{\emph{Results for Example 1 (a):} the difference $\Delta Q$ of the optimal success probabilities between the pure-mixed ($|\Psi_1\rangle$ and $\rho_2^0$) and pure-pure state ($|\Psi_1\rangle$ and $|\Psi_2^0\rangle$) discrimination as a function of the coherence for the finite dimensional state $|\Psi_2^0\rangle$ with binomial distributed probability amplitude. Dotted line for $N=10$ and solid line for $N=100$, where $P_1=0.15$, $t=0$ and $s_{1t'}=0.5$.}\label{fig4}
 \end{figure}

\begin{figure}
%[!htbp]
\centering
\includegraphics[width=8cm]{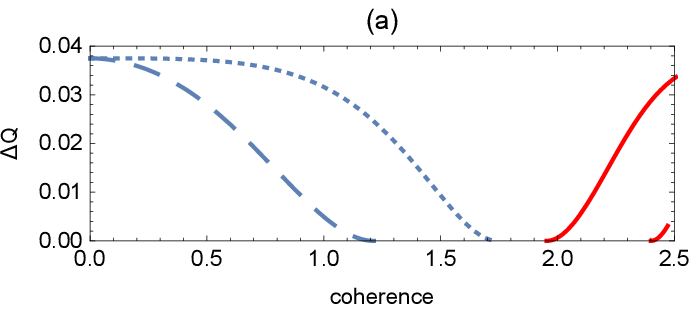} \\
\includegraphics[width=8cm]{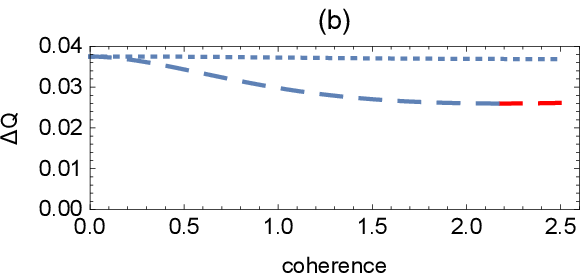} \\
 \caption{\emph{Results for Example 1 (b):} the difference $\Delta Q$ of the optimal success probabilities between the pure-mixed ($|\Psi_1\rangle$ and $\rho_2$) and pure-pure ($|\Psi_1\rangle$ and $|\Psi_2\rangle$) state discrimination as a function of the coherence for the infinite dimensional state $|\Psi_{2}\rangle$, with $P_1=0.15$ and $s_{1t'}=0.5$. Figs (a) and (b) correspond to the scheme with generalized well-known coherent and squeezed vacuum states, respectively. Solid line: $t=0$; dashed line: $t=3$; dotted line: $t=5$.}\label{fig5}
 \end{figure}

\begin{figure}
%[!htbp]
\centering
\includegraphics[width=8cm]{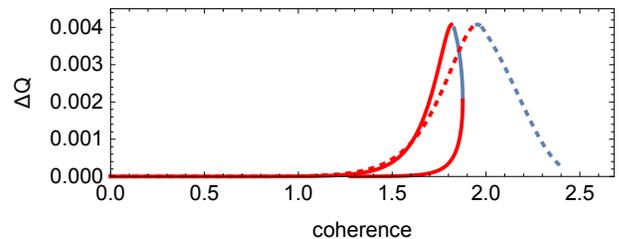} \\
 \caption{\emph{Results for Example 2 (a):} the difference $\Delta Q$ of the optimal success probabilities between the mixed-mixed ($\rho_1^0$ and $\rho_2^0$) and pure-pure ($|\Psi_1^0\rangle$ and $|\Psi_2^0\rangle$) state discrimination as a function of the coherence of $|\Psi_j^0\rangle$ ($j=1,2$) with $P_1=0.15$. Solid line: $N=10$, $s_{ii'}=0.5>\sqrt{\frac{P_1}{P_2}}$ ($i=1,...,4$), $s_{ii'}=0.2\leq\sqrt{\frac{P_1}{P_2}}$ ($i=5,...,10$); dotted line: $N=50$, $s_{ii'}=0.5>\sqrt{\frac{P_1}{P_2}}$ ($i=1,...,4$), $s_{ii'}=0.2\leq\sqrt{\frac{P_1}{P_2}}$ ($i=5,...,50$).
 }\label{fig6}
 \end{figure}

\begin{figure}
%[!htbp]
\centering
\includegraphics[width=8cm]{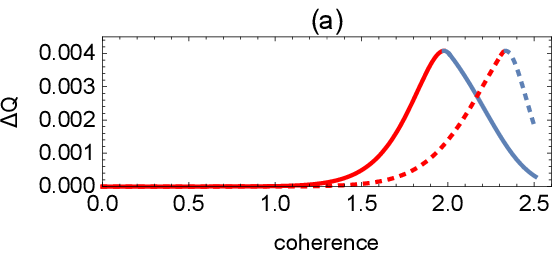} \\
\includegraphics[width=8cm]{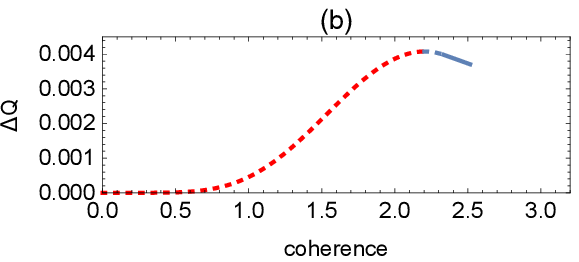} \\
 \caption{\emph{Results for Example 2 (b):}
 the difference $\Delta Q$ of the optimal success probabilities between the mixed-mixed ($\rho_1$ and $\rho_2$) and pure-pure ($|\Psi_1\rangle$ and $|\Psi_2\rangle$) state discrimination as functions of the coherence $C_{\rm{re}}(|\Psi_i\rangle)$.
 (\textbf{a}) and (\textbf{b}) correspond to the scheme with generalized well-known coherent and squeezed vacuum states, respectively, for $P_1=0.15$ and $P_2=0.85$. Solid line: $s_{ii'}=0.5>\sqrt{\frac{P_1}{P_2}}$ ($i=1,...,4$), $s_{ii'}=0.2\leq\sqrt{\frac{P_1}{P_2}}$ ($i=5,...,\infty$); dotted line: $s_{ii'}=0.2$ ($i=1,...,4$), $s_{ii'}=0.5$ ($i=5,...,\infty$).} \label{fig7}
\end{figure}

For the two examples above, the difference $\Delta Q$ of the optimal success probabilities  between the pure (mixed)-mixed and pure-pure state discrimination
is presented in Figs. \ref{fig4}-\ref{fig7}, corresponding to the binomial and Poisson distribution of photon numbers, respectively. It shows that the pure-pure state discrimination scheme is also superior to the pure (mixed)-mixed one and the coherence which is detrimental and helpful to state discriminations coexists irrespective of any schemes involved in the above two examples.

%尝试将图分开，有限维二项式态可以考虑单图单列。
%the coherence encoded in our superposed state which is larger than a critical value $C_c$ (for the calculations in Fig. \ref{fig3}, we have $C_c=1.11)$) is helpful for state discrimination ($\frac{d\Delta Q}{dC(|\Psi_1\rangle)}>0$); otherwise, it is detrimental to it ($\frac{d\Delta Q}{dC(|\Psi_1\rangle)}<0$). This is different from the results for finite dimensional space in Fig. \ref{fig2} where we can not find any helpful coherence. The numerical results corresponding to helpful coherence is all labeled as red lines in all of our paper.
%In addition, under the effect of this helpful coherence, the result of pure superposed state is superior to the one of mixed state without aid of the equal-fidelity condition given by Eq. (\ref{coherent state}).

For the quantum filtering including finite-dimensional systems associated with binomial distribution of photon numbers (shown in Fig. \ref{fig4}), the quantum coherence which is helpful to state discriminations can be acquired for larger $N$, which is not the case for the filtering with rank-two system in Fig. \ref{fig2}. Then, we make an extension to infinite dimensional systems corresponding to the generalized Gaussian states in Fig. \ref{fig5}.  As $N\to\infty$, the results of quantum filtering for mixed states (\ref{finite-dimensional mixed state}) and (\ref{infinite-dimensional mixed state}) gives rise to the same results since the binomial and Poisson distributions are equivalent to each other at this case. In addition, from the results in Fig. \ref{fig5}, it is indicated that as the parameter $t$ increases, the helpful coherence encoded in the well-known coherent state decreases by contrary. While for the scheme with the generalized squeezed vacuum state, despite the superiority of the pure-pure state scheme versus the pure-mixed one, the coherence contributes very little to this superiority, as is shown in Fig. \ref{fig5} (b).

For the results corresponding to mixed-mixed state discrimination schemes shown in Figures \ref{fig6} and \ref{fig7},
compared with the results of quantum filtering in Figures \ref{fig4} and \ref{fig5}, it can be concluded that the symmetrically (asymmetrically) distributed coherence is always helpful (detrimental) to state discrimination for lower dimensional systems. As the dimension increases, symmetrically (asymmetrically) distributed coherence may become detrimental (helpful) on the contrary. For $N\to\infty$, just like quantum filtering, the result for the binomial state is also equivalent to the one of the well-known coherent states for this mixed-mixed state discrimination scheme.

%which opposites to the result for finite systems in Sec. \ref{mixed local state discrimination1} and \ref{Simulation1}.
%As the dimension number increases, different from quantum filtering, coherence which is helpful to state discriminations is turned to be detrimental by contrary.
We also see that only a small range of helpful coherence is vital for state discrimination, while the others have little effect for the schemes with high-dimensional binomial and the generalized well-known coherent states, as shown in Figs. \ref{fig6} and \ref{fig7} (a) (solid line). In the cases including the generalized squeezed vacuum states, Fig. \ref{fig7} (b) shows that there are more regions of helpful coherence for the mixed-mixed scheme. Since the well-known coherent state is the eigenstate of the annihilation operator, it saturates the lower bound of the quantum uncertainty relation for momentum and position exactly ($\Delta p\Delta q=\hbar/2$). That is, the well-known coherent state approaches the boundary between the classical and quantum physics. Just because of this property, the coherence encoded in the infinite dimensional systems associated with this well-known state exhibits so many abnormal behaviors in unambiguous state discrimination, different from the results for finite dimensional systems and infinite ones associated with the squeezed vacuum states.

Concerning the related experiments in quantum optics, the discrimination of infinite dimensional quantum states such as the well-known coherent states is a subject of research significance \cite{Kennedy1973QPR,Banaszek1999PLA,Huttner1995PRA,Han2018NJP,Namkung2018SR1}. The phases in the well-known coherent state $|\alpha\rangle=e^{-|\alpha|^2/2}\sum_{i=0}^{\infty}\alpha^i/\sqrt{i!}|i\rangle$ ($\alpha=|\alpha|e^{{\rm{i}}\theta}$) are randomized under quantum decoherence. Then taking the average over the variable $\theta$, one has
$$\frac{1}{2\pi}\int_{0}^{2\pi}|\alpha\rangle\langle\alpha|d\theta=\sum_{i=0}^{\infty}\frac{1}{i!}|\alpha|^{2i}|i\rangle\langle i|.$$
Hence, the mixed state in (\ref{infinite-dimensional mixed state}) can be prepared successfully. Otherwise, the state can also be acquired via local measurements on a two-mode well-known coherent state.

\section{Summary and Outlook}\label{Summ}

We have investigated the discrimination between a pure state and a rank-$N$ mixed state (quantum filtering) and compared its optimal successful probability with the one for discriminating another two pure states. One state involved in the pure-pure scheme is identical to the one in quantum filtering; the other one is superposed by the eigenvectors of the above-mentioned mixed state. As the pure-mixed and pure-pure states have the same fidelity, we prove that the optimal success probability of a pure-pure state scheme is superior to quantum filtering. For lower dimensional systems, e.g., $N=2,3$, the coherence encoded in the pure state is detrimental to state discrimination. If the equal-fidelity restriction is relaxed and the phases in the constructed coherent pure states are identical to each other, the superiority of the pure-pure state scheme is impaired severely.  As we adjust the phases to proper values, the superiority of the pure-pure scheme revives, and helpful coherence is acquired. However, this superiority emerges not surprisingly because of a lower fidelity between the two pure states versus the pure-mixed one.

After discriminating two rank-$N$ ($N$ is a finite positive integer) mixed states whose eigenvectors have one-to-one non-zero overlaps (mixed-mixed state scheme), we also consider the discrimination of two pure states which are superposed by the eigenvectors. Thus, the pure-pure and mixed-mixed states also have the same fidelity. We also prove that the pure-pure state scheme is bound to be superior to the mixed-mixed one. Namely, the result of \emph{Theorem 1} %Is the italics necessary?
 in Ref. \cite{JHZhang2020PRA} confined to rank-two systems is generalized to rank-$N$ systems successfully. Due to the symmetrical distribution of coherence encoded in the two pure superposed states, different from the result of quantum filtering, the coherence is always helpful to state discrimination for lower-dimensional systems.

Finally, in order to generalize our results to infinite-dimensional systems, we have first
considered the examples of discriminating binomial states.
For higher dimensional systems, we remark that some asymmetrically (symmetrically) distributed coherence which is helpful (detrimental) to state discrimination occurs, which turns to be more apparent after we made an extension to infinite-dimensional systems ($N\to\infty$) associated with the well-known coherent rather than squeezed-vacuum states.
These results can be attributed to the fact that the well-known coherent state which saturates the lower bound of the quantum uncertainty relation for momentum and position approaches the boundary between classical and quantum physics.

Sequential state discrimination (SSD) provided in \cite{Bergou2013PRL} is a scheme for discriminating one sender's quantum states via $N$ observers who are separately located. SSD is investigated sequentially in \cite{Zhang2017arXiv,Namkung2018SR,JHZhang2020PRA,Namkung2020entropy}.
 As a next step, we plan to investigate another interesting problem corresponding to SSD including quantum filtering and rank-$N$ mixed states discriminations
 and consider the role played by quantum correlation and coherence in the procedure.

\begin{acknowledgments}
This work is supported by NSF of China (Grant Nos. 12171044, 12075159, 11675119), Shanxi Education Department Fund (2020L0543), Beijing Natural Science Foundation (Z190005), Academy for Multidisciplinary Studies, Capital Normal University, the Academician Innovation Platform of Hainan Province, and Shenzhen Institute for Quantum Science and Engineering, Southern University of Science and Technology (No. SIQSE202001).
\end{acknowledgments}

\end{document}